\documentclass[11pt]{article}

\usepackage{epsf}
\usepackage{amsmath}
\usepackage{amssymb}
\usepackage{graphicx}
\usepackage{dcolumn}
\usepackage{bm}
\usepackage{lscape}
\usepackage{tabularx}
\usepackage[titles]{tocloft}
\usepackage{color}

\newcommand{\append}[1]{\protect\refstepcounter{section}
                \section*{Appendix \thesection \, #1}
                \addcontentsline{toc}{section}{Appendix \thesection: #1}}

\begin{document}

\begin{center}
\Large\bf\boldmath
\vspace*{2.0cm}SuperIso Relic: A program for calculating relic density \\
and flavor physics observables in Supersymmetry
\unboldmath
\end{center}

\vspace{0.8cm}
\begin{center}
A. Arbey\footnote{Electronic address: \tt alexandre.arbey@ens-lyon.fr}\\[0.4cm]
{\sl Universit\'e de Lyon, Lyon, F-69000, France; Universit\'e Lyon~1,\\ Villeurbanne, F-69622, France;
Centre de Recherche Astrophysique de Lyon,\\ Observatoire de Lyon, 9 avenue Charles Andr\'e,
Saint-Genis Laval cedex,\\ F-69561, France; CNRS, UMR 5574; Ecole Normale Sup\'erieure de Lyon, Lyon, France.\\}
\vspace*{1cm}
F. Mahmoudi\footnote{Electronic address: \tt mahmoudi@in2p3.fr\\
\hspace*{0.48cm} URL: \tt http://superiso.in2p3.fr/relic}\\[0.4cm]
{\sl Laboratoire de Physique Corpusculaire de Clermont-Ferrand (LPC),\\ Universit\'e Blaise Pascal, CNRS/IN2P3, 63177 Aubi\`ere Cedex, France.}
\end{center}
\vspace{0.9cm}
\begin{abstract}
\noindent We describe \verb?SuperIso Relic?, a public program for evaluation of relic density and flavor physics observables in the minimal supersymmetric extension of the Standard Model (MSSM). \verb?SuperIso Relic? is an extension of the \verb?SuperIso? program which adds to the flavor observables of \verb?SuperIso? the computation of all possible annihilation and coannihilation processes of the LSP which are required for the relic density calculation. All amplitudes have been generated at the tree level with \verb?FeynArts?/\verb?FormCalc?, and widths of the Higgs bosons are computed with \verb?FeynHiggs? at the two-loop level. \verb?SuperIso Relic? also provides the possibility to modify the assumptions of the cosmological model, and to study their consequences on the relic density. 
\end{abstract}
\newpage
\tableofcontents
%
\newpage
\section{Introduction} 
The dark matter problem remains one of the most puzzling questions in cosmology. Cosmological analyses reveal that dark matter may be composed of non-baryonic particles, but their nature is still to be discovered. Many new physics models try to provide
a natural solution to the dark matter problem. Supersymmetry in particular offers a stable particle, the lightest supersymmetric particle (LSP), if $R$-parity is conserved, which could be the main component of the dark matter in the Universe. The current density of the LSP can be calculated and is referred as relic density. Compared to the latest precise WMAP measurements of the dark matter density \cite{wmap}, relic density can impose stringent constraints on the supersymmetric parameters.\\
\\
\verb?SuperIso Relic? is an extension of the \verb?SuperIso? program to the calculation of the relic density. The program calculates the relic density as well as the flavor physics observables using a SUSY Les Houches Accord file (SLHA1 \cite{slha1} or SLHA2 \cite{slha2}) as input, either generated automatically via a call to SOFTSUSY \cite{softsusy} or ISAJET \cite{isajet}, or provided by the user. The calculation can be performed automatically for different supersymmetry breaking scenarios, such as the minimal Supergravity Grand Unification (mSUGRA) also called Constrained MSSM (CMSSM), the Non-Universal Higgs Mass model (NUHM), the Anomaly Mediated Supersymmetry Breaking scenario (AMSB) and the Gauge Mediated Supersymmetry Breaking scenario (GMSB).\\
\\
One of the most important features of \verb?SuperIso Relic? in comparison to the other public relic density calculation codes, DarkSusy \cite{darksusy}, IsaRed \cite{isared} and Micromegas \cite{micromegas}, is that it provides the possibility to alter the underlying cosmological model, by modifying for example the radiation equation-of-state, the expansion rate or the thermal properties of the Universe in the period before Big-Bang nucleosynthesis (BBN), which is experimentally inaccessible and remains theoretically obscure. In \cite{arbey,arbey2}, it was shown that a modification of the expansion rate or of the entropy content of the Universe before BBN can strongly modify the calculated relic density and therefore change the relic density constraints on supersymmetric parameter space. \verb?SuperIso Relic? makes it possible to evaluate the uncertainties on the relic density due to the cosmological model, and inversely, to make prediction on the early Universe properties using the particle physics constraints.\\
\\
In the following, first the content of the \verb?SuperIso Relic? package will be presented, as well as the list of the main routines used for the relic density calculation. The procedure to use \verb?SuperIso Relic? will be then explained, and the inputs and outputs of the program will be introduced. Finally, some examples of results obtained with \verb?SuperIso Relic? will be given. In the Appendices, a description of the formulas and models used for computing the relic density will be detailed.
%
\section{Content of the SuperIso Relic package}%
\label{content}
\verb?SuperIso Relic? is a mixed C / Fortran program devoted to the calculation of the relic density and able to compute many flavor observables. Seven main programs are provided in the package, but the users are also invited to write their own main programs. In particular \verb?slha.c? can scan files written following the SUSY Les Houches Accord formats, and calculates the implemented observables. The main programs \verb?msugra.c?, \verb?amsb.c?, \verb?gmsb.c?, and \verb?nuhm.c? have to be linked to the \verb?SOFTSUSY? \cite{softsusy} and/or the \verb?ISASUGRA/ISAJET? \cite{isajet} packages, in order to compute supersymmetric mass spectra and couplings within respectively the mSUGRA, AMSB, GMSB or NUHM scenarios.\\
\\
The main steps to compute the observables in \verb?SuperIso Relic? are the following:
\begin{itemize}
\item Generation of a SLHA file with \verb?ISAJET? or \verb?SOFTSUSY? (or supply of a SLHA file by the user),\vspace*{-0.2cm}
\item Scan of the SLHA file,\vspace*{-0.2cm}
\item Calculation of the widths of the Higgs bosons with \verb?FeynHiggs?,\vspace*{-0.2cm}
\item Calculation of the squared amplitudes of the annihilation diagrams involved in the relic density calculation,\vspace*{-0.2cm}
\item Computation of the thermally averaged total annihilation cross section,\vspace*{-0.2cm}
\item Solving of the Boltzmann equations and computation of the relic density,\vspace*{-0.2cm}
\item Calculation of the flavor physics observables.
\end{itemize}
It should be noted that the relic density calculation is performed even if the LSP is a charged particle. A theoretical description of the calculation of the thermally averaged total annihilation cross section can be found in Appendix \ref{sigmav} and the detail of the calculation of the relic density in the cosmological standard model is given in Appendix \ref{costamo}. We refer to \cite{superiso} for a complete description of the calculation of the flavor observables.\\
\\
The processes involved in the relic density calculation are all the annihilation and co-annihilation processes of the type
\begin{equation}
  \tilde{i} + \tilde{j} \to k + l
\end{equation}
where $\tilde{i},\tilde{j}$ are supersymmetric particles and $k,l$ are Standard Model particles. The number of involved processes is more than 3000, and the number of diagrams is even larger. To generate all the squared amplitudes, we have written a \verb?Mathematica? \cite{mathematica} script which calls \verb?FeynArts? \cite{feynarts} and \verb?FormCalc? \cite{formcalc} and generates the necessary routines for the numerical computation of the amplitudes. These routines are part of \verb?SuperIso Relic? and can be found in \verb?src/relic?. They rely on the \verb?LoopTools? \cite{formcalc} package to initialize \verb?FormCalc? internal routines, as well as \verb?FeynHiggs? \cite{feynhiggs} to calculate the widths of the Higgs bosons which is performed at two-loop level.\\
\verb?LoopTools 2.4? and \verb?FeynHiggs 2.6.5? are included in the \verb?SuperIso Relic v2.5? package, and they can be found in \verb?src/contrib?. Therefore the user does not need to download these programs separately.\\
\\
The compilation process of all the needed routines is very long (a few hours), and their calculation can take time. Fortunately, all the squared amplitude routines are not necessary at the same time, as some processes have only negligible effects. Therefore, all the squared amplitudes are not computed for a SUSY parameter space point, and a selection is performed to save time, as described in Appendix \ref{sigmav}.

\subsection{Parameter structures}
The package \verb?SuperIso Relic? relies on the definition of a main structure in \verb?src/include.h?, which is defined as follows:\\
\begin{verbatim}
typedef struct parameters
/* structure containing all the scanned parameters from the SLHA file */
{
	int model; /* mSUGRA = 1, GMSB = 2, AMSB = 3 */
	int generator; /* ISAJET = 1, SOFTSUSY = 2 */
	double Q; /* Qmax ; default = M_EWSB = sqrt(m_stop1*mstop2) */
	double m0,m12,tan_beta,sign_mu,A0; /* mSUGRA parameters */
	double Lambda,Mmess,N5,cgrav,m32; /* AMSB, GMSB parameters */
	double mass_Z,mass_W,mass_b,mass_top_pole,mass_tau_pole; /* SM parameters */
	double inv_alpha_em,alphas_MZ,alpha,Gfermi,GAUGE_Q; /* SM parameters */
	double charg_Umix[3][3],charg_Vmix[3][3],stop_mix[3][3],sbot_mix[3][3],
	stau_mix[3][3],neut_mix[6][6],mass_neut[6]; /* mass mixing matrices */
	double Min,M1_Min,M2_Min,M3_Min,At_Min,Ab_Min,Atau_Min,M2H1_Min,M2H2_Min,
	mu_Min,M2A_Min,tb_Min,mA_Min; /* optional input parameters at scale Min */
	double MeL_Min,MmuL_Min,MtauL_Min,MeR_Min,MmuR_Min,
	MtauR_Min; /* optional input parameters at scale Min */
	double MqL1_Min,MqL2_Min,MqL3_Min,MuR_Min,McR_Min,MtR_Min,
	MdR_Min,MsR_Min,MbR_Min; /* optional input parameters at scale Min */
	double N51,N52,N53,M2H1_Q,M2H2_Q; /* optional input parameters (N51...3: GMSB)  */
	double mass_d,mass_u,mass_s,mass_c,mass_t,mass_e,mass_nue,mass_mu,
	mass_num,mass_tau,mass_nut; /* SM masses */
	double mass_gluon,mass_photon,mass_Z0; /* SM masses */
	double mass_h0,mass_H0,mass_A0,mass_H,mass_dnl,mass_upl,mass_stl,mass_chl,
	mass_b1,mass_t1; /* Higgs & superparticle masses */
	double mass_el,mass_nuel,mass_mul,mass_numl,mass_tau1,mass_nutl,mass_gluino,
	mass_cha1,mass_cha2; /* superparticle masses */
	double mass_dnr,mass_upr,mass_str,mass_chr,mass_b2,mass_t2,mass_er,mass_mur,
	mass_tau2; /* superparticle masses */
	double mass_nuer,mass_numr,mass_nutr,mass_graviton,
	mass_gravitino; /* superparticle masses */
	double gp,g2,g3,YU_Q,yut[4],YD_Q,yub[4],YE_Q,yutau[4]; /* Yukawa couplings */
	double HMIX_Q,mu_Q,tanb_GUT,Higgs_VEV,mA2_Q,MSOFT_Q,M1_Q,M2_Q,
	M3_Q; /* parameters at scale Q */
	double MeL_Q,MmuL_Q,MtauL_Q,MeR_Q,MmuR_Q,MtauR_Q,MqL1_Q,MqL2_Q,MqL3_Q,MuR_Q,McR_Q,
	MtR_Q,MdR_Q,MsR_Q,MbR_Q; /* masses at scale Q */
	double AU_Q,A_u,A_c,A_t,AD_Q,A_d,A_s,A_b,AE_Q,A_e,A_mu,A_tau; /* trilinear couplings */
	
	/* SLHA2 */
	int NMSSM,Rparity,CPviolation,Flavor;
	double mass_nutau2,mass_e2,mass_nue2,mass_mu2,mass_numu2,mass_d2,mass_u2,
	mass_s2,mass_c2;
	double CKM_lambda,CKM_A,CKM_rho,CKM_eta;
	double PMNS_theta12,PMNS_theta23,PMNS_theta13,PMNS_delta13,PMNS_alpha1,PMNS_alpha2;
	double lambdaNMSSM_Min,kappaNMSSM_Min,AlambdaNMSSM_Min,AkappaNMSSM_Min,
	lambdaSNMSSM_Min,xiFNMSSM_Min,xiSNMSSM_Min,mupNMSSM_Min,mSp2NMSSM_Min,mS2NMSSM_Min,
	mass_H03,mass_A02,NMSSMRUN_Q,lambdaNMSSM,kappaNMSSM,AlambdaNMSSM,AkappaNMSSM,
	lambdaSNMSSM,xiFNMSSM,xiSNMSSM,mupNMSSM,mSp2NMSSM,mS2NMSSM; /* NMSSM parameters */
	double PMNSU_Q,CKM_Q,MSE2_Q,MSU2_Q,MSD2_Q,MSL2_Q,MSQ2_Q,TU_Q,TD_Q,TE_Q;
	
	double CKM[4][4]; /* CKM matrix */
	double H0_mix[4][4],A0_mix[4][4]; /* Higgs mixing matrices */
	double sU_mix[7][7],sD_mix[7][7],sE_mix[7][7], sNU_mix[4][4]; /* mixing matrices */
	double sCKM_msq2[4][4],sCKM_msl2[4][4],sCKM_msd2[4][4],sCKM_msu2[4][4],
	sCKM_mse2[4][4]; /* super CKM matrices */
	double PMNS_U[4][4]; /* PMNS mixing matrices */
	double TU[4][4],TD[4][4],TE[4][4]; /* trilinear couplings */
	
	/* non-SLHA*/
	double mass_b_1S,mass_b_pole,mtmt;
	double Lambda5; /* Lambda QCD */
	
	/* Flavor constants */
	double f_B,f_Bs,f_Ds,m_B,m_Bs,m_Ds,m_K,m_Kstar,m_D,life_B,life_Bs,life_Ds;
	
	/* Decay widths */
	double width_h0,width_H0,width_A0,width_H;
}
parameters;
\end{verbatim}
This structure contains all the important parameters and is called by most of the main functions in the program. An additional structure specific to the relic density calculation is also defined:
\begin{verbatim}
typedef struct relicparam
/* structure containing the cosmological model parameters */
{                                                           
        double dd0,ndd;                                     
        double sd0,nsd;                                     
        double table_eff[276][3];
}
relicparam;
\end{verbatim}
This structure is used to define the cosmological model based on which the relic density calculation is performed.
\subsection{Main routines}
\label{mainroutines}
We now review the main routines of the code needed for the relic density calculation. For the main procedures related to the flavor observable calculations we refer the reader to \cite{superiso}.\\
\\
The most relevant C routines are the following:
\begin{itemize}
\item \verb?void Init_param(struct parameters* param)?\\
\\
This function initializes the \verb?param? structure, setting all the parameters to 0, apart from the SM masses and the value of the strong coupling constant at the $Z$-boson mass, which receive the values given in the PDG08 \cite{PDG}.\\

\item \verb?int Les_Houches_Reader(char name[], struct parameters* param)?\\
\\
This routine reads the SLHA file named \verb?name?, and put all the read parameters in the structure \verb?param?. It should be noted that a negative value for \verb?param->model? indicates a problem in reading the SLHA file, or a model not yet included in SuperIso (such as $R$-parity breaking models). In this case, \verb?Les_Houches_Reader? returns 0, otherwise 1.\\

\item \verb?int test_slha(char name[])?\\
\\
This routine checks if the SLHA file is valid, and if so returns 1. If not, -1 means that in the SLHA generator the computation did not succeed ({\it e.g.} because of tachyonic particles), -2 means that the considered model is not currently implemented in \verb?SuperIso?, and -3 indicates that the provided file is either not in the SLHA format, or some important elements are missing.\\
\item \verb?int softsusy_sugra(double m0, double m12, double tanb, double A0,?\\
\verb?double sgnmu, double mtop, double mbot, double alphas_mz, char name[])?
\item \verb?int isajet_sugra(double m0, double m12, double tanb, double A0, ?\\
\verb?double sgnmu, double mtop, char name[])?
\item \verb?int softsusy_gmsb(double Lambda, double Mmess, double tanb, int N5,?\\
\verb? double cGrav, double sgnmu, double mtop, double mbot, double alphas_mz,?\\
\verb? char name[])?
\item \verb?int softsusy_amsb(double m0, double m32, double tanb, double sgnmu,?\\
\verb?double mtop, double mbot, double alphas_mz, char name[])?
\item \verb?int softsusy_nuhm(double m0, double m12, double tanb, double A0, double mu,?\\ 
\verb?double mA, double mtop, double mbot, double alphas_mz, char name[])?\\
\\
The above routines call \verb?SOFTSUSY? or \verb?ISAJET? to compute the mass spectrum corresponding to the input parameters (more details are given in the next sections), and return a SLHA file whose name has to be specified in the string \verb?name?.\\
\\
\item \verb?void ModelIni(struct parameters* param, double relicmass, double maxenergy)?\\
\\
This routine is an interface between the C routines and the Fortran routines and it defines all the Fortran variables using the C variables.\\[0.1cm]
\item \verb?double findrelicmass(struct parameters* param, int *scalar)?\\                
\\
This function determines the LSP mass, and checks if the LSP is scalar (\verb?*scalar=1?) or fermionic (\verb?*scalar=0?).\\[0.1cm]
\item \verb?int Weff(double* res, double sqrtS, struct parameters* param, double relicmass)?\\
\\
This function calls the Fortran routines and returns the effective annihilation rate $W_{\rm{eff}}$ at a given center of mass energy \verb?sqrtS?, following the procedure described in Appendix \ref{sigmav}.\\[0.0cm]
\item \verb?int Init_relic(double Wefftab[NMAX][2], int *nlines_Weff, struct parameters* param)?\\
\\
This routine computes for different values of $\sqrt{s}$ the effective annihilation rates $W_{\rm{eff}}$ needed for the calculation of $\langle \sigma v \rangle$ using the \verb?Weff? function, and collects them in table \verb?Wefftab?.\\[0.1cm]
\item \verb?double sigmav(double T, double relicmass, double Wefftab[NMAX][2], int nlines,?\\
\verb?struct parameters* param)?\\
\\
This function computes the averaged annihilation cross section $\langle \sigma v\rangle$ using the effective annihilation rates $W_{\rm{eff}}$ collected in table \verb?Wefftab?.\\[0.1cm]
\item \verb?double heff(double T, struct relicparam* paramrelic)?\\
\verb?double sgstar(double T, struct relicparam* paramrelic)?\\
\verb?double geff(double T, struct relicparam* paramrelic)?\\
\\
These three functions compute respectively $h_{\rm{eff}}$, $\sqrt{g_*}$ and $g_{\rm{eff}}$ at the temperature~\verb?T?.\\[0.1cm]
\item \verb?double Yeq(double T,struct parameters* param, struct relicparam* paramrelic)?\\
\verb?double dYeq_dT(double T,struct parameters* param, struct relicparam* paramrelic)?\\
\\
The first function computes $Y_{eq}$ at a temperature \verb?T?, and the second one its derivative.\\[0.1cm]
\item \verb?double Tfo(double Wefftab[NMAX][2], int nlines_Weff, double relicmass,?\\
\verb?struct parameters* param, double d, struct relicparam* paramrelic)?\\
\\
This function computes the freeze-out temperature following Eq. (\ref{Tfo}), using the \verb?Wefftab? generated previously.\\[0.1cm]
\item \verb?double relic_density(double Wefftab[NMAX][2], int nlines_Weff,?\\
\verb?struct parameters* param, struct relicparam* paramrelic)?\\
\verb?double relic_calculator(char name[])?\\
\\
This main procedure computes the relic density using the \verb?Wefftab? generated previously. \verb?relic_calculator? is a container function which scans the SLHA file and computes the relic density.\\[0.1cm]
\item \verb?void Init_cosmomodel(struct relicparam* paramrelic)?\\
\verb?void Init_modeleff(int model_eff, struct relicparam* paramrelic)?\\
\verb?void Init_dark_density(double dd0, double ndd, struct relicparam* paramrelic)?\\
\verb?void Init_dark_entropy(double sd0, double nsd, struct relicparam* paramrelic)?\\
\\
These procedures define the cosmological model based on which the relic density is computed. \verb?Init_cosmomodel? has to be called first to initialize the \verb?paramrelic? structure. To alter the QCD equation-of-state as in Appendix \ref{qcdstate}, \verb?Init_modeleff? must be called while specifying the model: \verb?model_eff?$=1\cdots5$ corresponds respectively to the models A, B, B2, B3 and C developed in \cite{hindmarsh05}, and \verb?model_eff?$=0$ to the old model formerly used in \verb?Micromegas? and \verb?DarkSusy?. If not specified, the model is set by default to B (\verb?model_eff?$=2$). \verb?Init_dark_density? adds a dark energy density as in Eq. (\ref{rhoD}), with \verb?dd0?=$\kappa_\rho$ and \verb?ndd?=$n_\rho$, and \verb?Init_dark_entropy? adds a dark entropy density as in Eq. (\ref{sD}), with \verb?sd0?=$\kappa_s$ and \verb?nsd?=$n_s$. If these routines are not called, no additional density will be added, and the calculation will be performed in the standard cosmological model.\\[0.1cm]
\item \verb?double dark_density(double T, struct relicparam* paramrelic)?\\
\verb?double dark_entropy(double T, struct relicparam* paramrelic)?\\
\verb?double dark_entropy_derivative(double T, struct relicparam* paramrelic)?\\
\verb?double dark_entropy_Gd(double T, struct relicparam* relicparam)?\\
\\
These functions compute energy and entropy densities needed for the alternative cosmological models described in Appendix \ref{mocomo}.\\[0.1cm]
\item \verb?int FeynHiggs(char name[], struct parameters* param)?\\
\\
This routine calls \verb?FeynHiggs? to compute the widths and masses of the Higgs bosons corresponding to the SLHA file \verb?name? at the two-loop level, and puts these variables in the \verb?param? structure.\\
\end{itemize}
The complete list of C procedures implemented in \verb?SuperIso Relic? is available in \verb?src/include.h?.\\
\\
The Fortran routines can be found in \verb?src/relic?. They have been generated automatically by a \verb?Mathematica?/\verb?FormCalc? script and they perform the computation of all squared amplitudes. Because of the large number of these routines they will not be described further here. For the \verb?FormCalc? specific routines, we refer the reader to the \verb?FormCalc? manual \cite{formcalc}.
%
\section{Compilation and installation instructions}%
\label{compilation}
The \verb?SuperIso Relic? package can be downloaded from:\\
\\
{\tt http://superiso.in2p3.fr/relic}\\
\\
and is available in two versions:
\begin{itemize}
\item the shared library based version, \verb?SuperIso Relic Shared?, compiles the squared amplitude procedures on-the-fly, if they are needed. The compilation of this version is fast, but its running is slow. It is mostly intended for calculation of test-points.\\
\item the static library based version, \verb?SuperIso Relic?. Here all the squared amplitude procedures need to be compiled before running, and therefore the compilation process can take a couple of hours. This version computes much faster than the shared library version, and is therefore intended for the calculation of a large number of SUSY points.
\end{itemize}
The following main directory is created after unpacking:\\
\\
\verb?superiso_vX.X?\\
\\
This directory contains the \verb?src/? directory, in which all the source files can be found. The main directory contains also a \verb?Makefile?, a \verb?README?, seven sample main programs (\verb?msugra.c?, \verb?amsb.c?, \verb?gmsb.c?, \verb?nuhm.c?, \verb?slha.c?, \verb?test_modeleff.c? and \verb?test_standmod.c?) and one example of input file in the SUSY Les Houches Accord format (\verb?example.lha?). The compilation options should be defined in the \verb?Makefile?, as well as the path to the ISAJET \verb?isasugra.x? and SOFTSUSY \verb?softpoint.x? executable files, when needed. The important compilation options to be set are the C compiler name (by default \verb?gcc?) and the Fortran compiler name (by default \verb?gfortran?), and their specific flags if needed. Instructions are given in \verb?Makefile? for the Intel compilers and for the Fortran compiler \verb?g77?. For the \verb?SuperIso Relic Shared? version the \verb?LD? linker name (by default \verb?ld?) and its specific flags must also be set.\\
\verb?SuperIso Relic? is written for a C compiler respecting the C99 standard and a Fortran compiler. In particular, it has been tested successfully with the GNU C and GNU Fortran Compilers and the Intel C and Intel Fortran Compilers on Linux and Mac 32-bits or 64-bits machines, and with the latest versions of \verb?SOFTSUSY? and \verb?ISAJET?. Additional information can be found in the \verb?README? file.\\
To compile the library, type\\
\\
\verb?make?\\
\\
\noindent This creates \verb?libisospin.a? in \verb?src/? and \verb?librelic.a? in \verb?src/relic?, and compiles \verb?FeynHiggs? and \verb?LoopTools?. Then, to compile one of the seven programs provided in the main directory, type\\
\\
\verb?make name    ? or \verb?    make name.c?\\
\\
\noindent where \verb?name? can be \verb?msugra?, \verb?amsb?, \verb?gmsb?, \verb?slha?, \verb?nuhm?, \verb?test_modeleff? or \verb?test_standmod?. This generates an executable program with the \verb?.x? extension. Note that \verb?slha?, \verb?test_modeleff? and \verb?test_standmod? do not need \verb?ISAJET? or \verb?SOFTSUSY? programs, but \verb?msugra?, \verb?amsb?, \verb?gmsb? and \verb?nuhm? need at least one of them.\\
\\
\verb?slha.x? calculates the implemented observables, using the parameters contained in the SLHA file whose name has to be passed as input parameter.\\
\\
\verb?amsb.x?, \verb?gmsb.x?, \verb?msugra.x? and \verb?nuhm.x? compute the observables, starting first by calculating the mass spectrum and couplings thanks to \verb?ISAJET? (for \verb?msugra.x? only) and/or \verb?SOFTSUSY? within respectively the AMSB, GMSB, mSUGRA or NUHM parameter spaces.\\
\\
\verb?test_modeleff? and \verb?test_standmod? calculate the relic density, using the parameters contained in the SLHA file whose name has to be passed as input parameter, in the cosmological models described in the Appendices.\\
%
\section{Input and output description}%
\label{sample}
The input and output of the main programs are described in the following.

\subsection{mSUGRA inputs}
The program \verb?msugra.x? computes the observables in the mSUGRA parameter space, using \verb?ISAJET? and/or \verb?SOFTSUSY? to generate the mass spectra. If only one of these generators is available the corresponding \verb?#define? in \verb?msugra.c? has to be commented. The necessary arguments to this program are:
\begin{itemize}
 \item $m_0$: universal scalar mass at GUT scale,\vspace*{-0.2cm}
 \item $m_{1/2}$: universal gaugino mass at GUT scale,\vspace*{-0.2cm}
 \item $A_0$: trilinear soft breaking parameter at GUT scale,\vspace*{-0.2cm}
 \item $\tan\beta$: ratio of the two Higgs vacuum expectation values.
\end{itemize}
Optional arguments can also be given:
\begin{itemize}
 \item $sign(\mu)$: sign of Higgsino mass term, positive by default,\vspace*{-0.2cm}
 \item $m_t^{pole}$: top quark pole mass, by default 172.4 GeV,\vspace*{-0.2cm}
 \item $\overline{m}_b(\overline{m}_b)$: scale independent b-quark mass, by default 4.2 GeV (option only available for \verb?SOFTSUSY?),\vspace*{-0.2cm}
 \item $\alpha_s(M_Z)$: strong coupling constant at the $Z$-boson mass, by default 0.1176 (option only available for \verb?SOFTSUSY?).
\end{itemize}
If the arguments are not specified, a message will describe the needed parameters in the correct order.\\
\\
With \verb?SOFTSUSY? 2.0.18 and \verb?ISAJET? 7.78, running the program with:
\begin{verbatim}
./msugra.x 1000 1000 0 10
\end{verbatim}
returns
\begin{verbatim}
SLHA file generated by SOFTSUSY                                                                                      
delta0=8.116e-02                                                                                                     
BR_bsgamma=3.082e-04                                                                                                 
BR_Btaunu=1.100e-04                                                                                                  
Rtaunu=9.980e-01                                                                                                     
BR_Kmunu/BR_pimunu=6.454e-01                                                                                         
Rl23=1.000e+00                                                                                                       
BR_BDtaunu=6.976e-03                                                                                                 
BR_BDtaunu/BR_BDenu=2.974e-01
BR_Bsmumu=3.095e-09
BR_Dstaunu=4.818e-02
BR_Dsmunu=4.975e-03
a_muon=9.756e-11
charged_LSP=0
excluded_mass=0
omega=9.087e+00

SLHA file generated by ISAJET
delta0=8.094e-02
BR_bsgamma=3.072e-04
BR_Btaunu=1.099e-04
Rtaunu=9.980e-01
BR_Kmunu/BR_pimunu=6.454e-01
Rl23=1.000e+00
BR_BDtaunu=6.968e-03
BR_BDtaunu/BR_BDenu=2.974e-01
BR_Bsmumu=3.468e-09
BR_Dstaunu=4.813e-02
BR_Dsmunu=4.970e-03
a_muon=1.013e-10
charged_LSP=0
excluded_mass=0
omega=9.102e+00

\end{verbatim}
where \verb?delta0? refers to the isospin symmetry breaking in $B \to K^* \gamma$ decays, \verb?BR_bsgamma? the branching ratio of $B \to X_s \gamma$, \verb?BR_Btaunu? the branching ratio of $B_u \to \tau \nu_\tau$, \verb?Rtaunu? the normalized ratio to the SM value,  \verb?BR_Kmunu/BR_pimunu? the ratio ${\rm BR}(K \to \mu \nu_\mu)/{\rm BR}(\pi \to \mu \nu_\mu)$, \verb?Rl23? the ratio $R_{\ell 23}$, \verb?BR_BDtaunu? the branching ratio of $B \to D^0 \tau \nu_\tau$, \verb?BR_BDtaunu/BR_BDenu? the ratio ${\rm BR}(B \to D^0 \tau \nu_\tau)/{\rm BR}(B \to D^0 e \nu_e)$, \verb?BR_Bsmumu? the branching ratio of $B_s \to \mu^+ \mu^-$, \verb?BR_Dstaunu? and \verb?BR_Dsmunu? the branching ratios of $D_s \to \tau \nu_\tau$ and $D_s \to \mu \nu_\mu$ respectively, \verb?a_muon? the deviation in the anomalous magnetic moment of the muon, \verb?charged_LSP? indicates that the LSP is charged if equal to 1, \verb?excluded_mass? that the point is already excluded by the direct searches if it is equal to 1, and \verb?omega? is the relic density $\Omega h^2$. More details on the definitions and calculations of the flavor observables are given in \cite{superiso}.

\subsection{AMSB inputs}
The program \verb?amsb.x? computes the observables using the corresponding parameters generated by \verb?SOFTSUSY? in the AMSB scenario. The necessary arguments to this program are:
\begin{itemize}
 \item $m_0$: universal scalar mass at GUT scale,\vspace*{-0.2cm}
 \item $m_{3/2}$: gravitino mass at GUT scale,\vspace*{-0.2cm}
 \item $\tan\beta$: ratio of the two Higgs vacuum expectation values.
\end{itemize}
Optional arguments are the same as for mSUGRA. If the input parameters are missing, a message will ask for them.\\
\\
With \verb?SOFTSUSY? 2.0.18, running the program with:
\begin{verbatim}
./amsb.x 500 5000 5 -1
\end{verbatim}
returns
\begin{verbatim}
delta0=7.669e-02
BR_bsgamma=3.296e-04
BR_Btaunu=1.097e-04
Rtaunu=9.953e-01
BR_Kmunu/BR_pimunu=6.454e-01
Rl23=1.000e+00
BR_BDtaunu=6.971e-03
BR_BDtaunu/BR_BDenu=2.972e-01
BR_Bsmumu=3.349e-09
BR_Dstaunu=4.818e-02
BR_Dsmunu=4.975e-03
a_muon=-6.469e-10
excluded_mass=1
omega=1.291e-04
\end{verbatim}

\subsection{GMSB inputs}
The program \verb?gmsb.x? computes the observables using the GMSB parameters generated by \verb?SOFTSUSY?. The necessary arguments to this program are:
\begin{itemize}
 \item $\Lambda$: scale of the SUSY breaking in GeV (usually 10000-100000 GeV),\vspace*{-0.2cm}
 \item $M_{mess}$: messenger mass scale ($> \Lambda$),\vspace*{-0.2cm}
 \item $N_5$: equivalent number of $5+\bar{5}$ messenger fields,\vspace*{-0.2cm}
 \item $\tan\beta$: ratio of the two Higgs vacuum expectation values.
\end{itemize}
Optional arguments are the same as for mSUGRA, with an additional one:
\begin{itemize}
 \item $c_{Grav}$ ($\ge 1$): ratio of the gravitino mass to its value for a breaking scale $\Lambda$, 1 by default.
\end{itemize}
Again, in the case of lack of arguments, a message will be displayed.\\
\\
With \verb?SOFTSUSY? 2.0.18, running the program with:
\begin{verbatim}
./gmsb.x 2e4 5e6 1 10
\end{verbatim}
returns
\begin{verbatim}
delta0=6.309e-02
BR_bsgamma=4.285e-04
BR_Btaunu=8.078e-05
Rtaunu=7.331e-01
BR_Kmunu/BR_pimunu=6.439e-01
Rl23=9.988e-01
BR_BDtaunu=6.542e-03
BR_BDtaunu/BR_BDenu=2.789e-01
BR_Bsmumu=3.802e-09
BR_Dstaunu=4.806e-02
BR_Dsmunu=4.962e-03
a_muon=3.778e-08
excluded_mass=1
omega=3.497e-02
\end{verbatim}

\subsection{NUHM inputs}
The program \verb?nuhm.x? computes the observables using the NUHM parameters generated by \verb?SOFTSUSY?. The necessary arguments to this program are the same as for mSUGRA, with two additional ones, the values of $\mu$ and $m_A$:
\begin{itemize}
 \item $m_0$: universal scalar mass at GUT scale,\vspace*{-0.2cm}
 \item $m_{1/2}$: universal gaugino mass at GUT scale,\vspace*{-0.2cm}
 \item $A_0$: trilinear soft breaking parameter at GUT scale,\vspace*{-0.2cm}
 \item $\tan\beta$: ratio of the two Higgs vacuum expectation values,\vspace*{-0.2cm}
 \item $\mu$: $\mu$ parameter,\vspace*{-0.2cm}
 \item $m_A$: CP-odd Higgs mass.
\end{itemize}
Optional arguments can also be given:
\begin{itemize}
 \item $m_t^{pole}$: top quark pole mass, by default 172.4 GeV,\vspace*{-0.2cm}
 \item $\overline{m}_b(\overline{m}_b)$: scale independent b-quark mass, by default 4.2 GeV,\vspace*{-0.2cm}
 \item $\alpha_s(M_Z)$: strong coupling constant at the $Z$-boson mass, by default 0.1176.
\end{itemize}
In the absence of arguments, a message will be shown.\\
\\
With \verb?SOFTSUSY? 2.0.18, running the program with:
\begin{verbatim}
./nuhm.x 500 500 0 50 500 500
\end{verbatim}
returns
\begin{verbatim}
delta0=1.069e-01
BR_bsgamma=2.006e-04
BR_Btaunu=6.775e-05
Rtaunu=6.148e-01
BR_Kmunu/BR_pimunu=6.430e-01
Rl23=9.982e-01
BR_BDtaunu=6.333e-03
BR_BDtaunu/BR_BDenu=2.700e-01
BR_Bsmumu=3.338e-08
BR_Dstaunu=4.796e-02
BR_Dsmunu=4.952e-03
a_muon=2.188e-09
excluded_mass=0
omega=7.533e-02
\end{verbatim}
\subsection{SLHA input file}
The program \verb?slha.x? calculates the observables while reading the needed parameters in a given SLHA file. For example, the command
\begin{verbatim}
./slha.x example.lha
\end{verbatim}
returns
\begin{verbatim}
delta0=8.259e-02
BR_bsgamma=2.974e-04
BR_Btaunu=1.097e-04
Rtaunu=9.966e-01
BR_Kmunu/BR_pimunu=6.454e-01
Rl23=1.000e+00
BR_BDtaunu=6.966e-03
BR_BDtaunu/BR_BDenu=2.973e-01
BR_Bsmumu=3.470e-09
BR_Dstaunu=4.813e-02
BR_Dsmunu=4.969e-03
a_muon=1.938e-10
excluded_mass=0
omega=1.188e+01
\end{verbatim}
If the SLHA file provided to \verb?slha.x? is invalid, a message will be displayed:
\begin{itemize}
\item \verb?Invalid point? means that the SLHA generator had not succeeded in generating the mass spectrum ({\it e.g.} due to the presence of tachyonic particles).\vspace*{-0.2cm}
\item \verb?Model not yet implemented? means that the SLHA file is intended for a model not implemented in SuperIso, such as $R$-parity violating models.\vspace*{-0.2cm}
\item \verb?Invalid SLHA file? means that the SLHA file is broken and misses important parameters.
\end{itemize}
\subsection{Alternative QCD equations of state}
The program \verb?test_modeleff.x? calculates the relic density while reading the needed parameters in the SLHA file, for the different QCD equations of state ({\it i.e.} alternative models of $g_{\rm{eff}}$ and $h_{\rm{eff}}$) described in Appendix \ref{qcdstate}. For example, the command
\begin{verbatim}
./test_modeleff.x example.lha
\end{verbatim}
returns
\begin{verbatim}
Dependence of the relic density on the calculation of heff and geff
For model_eff=1 (model A): omega=1.188e+01
For model_eff=2 (model B (default)): omega=1.188e+01
For model_eff=3 (model B2): omega=1.196e+01
For model_eff=4 (model B3): omega=1.181e+01
For model_eff=5 (model C): omega=1.189e+01
For model_eff=0 (old model): omega=1.165e+01
\end{verbatim}
\subsection{Effective energy and entropy densities}
The program \verb?test_standmodel.x? reads the needed parameters in the SLHA file, and calculates the relic density while adding to the standard cosmological model an effective energy density such that
\begin{equation}
 \rho_D =  \kappa_\rho \rho_{rad}(T_{BBN}) \bigl(T/T_{BBN}\bigr)^{n_\rho} \;,
\end{equation}
and/or an effective entropy density
\begin{equation}
 s_D =  \kappa_s s_{rad}(T_{BBN}) \bigl(T/T_{BBN}\bigr)^{n_s} \;,
\end{equation}
which modify the Early Universe properties without having observational consequences if chosen adequately \cite{arbey2}. A description of the model and of the related equations can be found in Appendix \ref{mocomo}. The necessary arguments to this program are\footnote{The preferential values given inside the brackets correspond to cosmological models without observational consequences, {\it i.e.} as valid as the cosmological standard model.}:
\begin{itemize}
 \item SLHA file name,\vspace*{-0.2cm}
 \item $\kappa_\rho$: ratio of dark energy density over radiation energy density at BBN time (preferentially $<1$),\vspace*{-0.2cm}
 \item $n_\rho$: dark energy density decrease exponent (preferentially $>4$),\vspace*{-0.2cm}
 \item $\kappa_s$: ratio of dark entropy density over radiation entropy density at BBN time (preferentially $<1$),\vspace*{-0.2cm}
 \item $n_s$: dark entropy density decrease exponent (preferentially $>3$).
\end{itemize}
Note that $n_\rho=4$ corresponds to a radiation-like energy density, $n_\rho=6$ to a quintessence-like energy density and $n_\rho=8$ to a decaying scalar field energy density. Also, $n_s=3$ corresponds to a radiation-like entropy density and $n_s=4$ can appear in reheating models.\\
For example, the command
\begin{verbatim}
./test_standmod.x example.lha 1e-3 6 1e-3 4
\end{verbatim}
returns
\begin{verbatim}
For the cosmological standard model:
omega=1.188e+01
For the specified model with dark density/entropy:
omega=4.633e+02
\end{verbatim}
\noindent Using the aforementioned main programs as examples, the user is encouraged to write his/her own programs in order to, for example, perform scans in a given supersymmetric scenario, or test different cosmological models.\\
\section{Results}
\label{result}
\begin{figure}[t!]
\begin{center}
\includegraphics[angle=270,width=10cm]{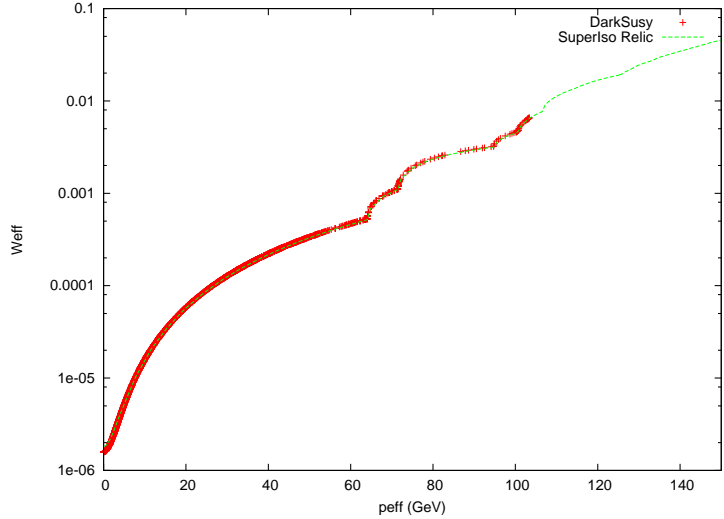}
\end{center}
\caption{$W_{\rm eff}$ in function of $p_{\rm eff}$, computed with SuperIso Relic (dashed green line), and with DarkSusy (red crosses). This comparison shows an excellent agreement.\label{WeffDS}}
\end{figure}
\begin{figure}[t!]
\begin{center}
\hspace*{0.2cm}\includegraphics[width=7.5cm]{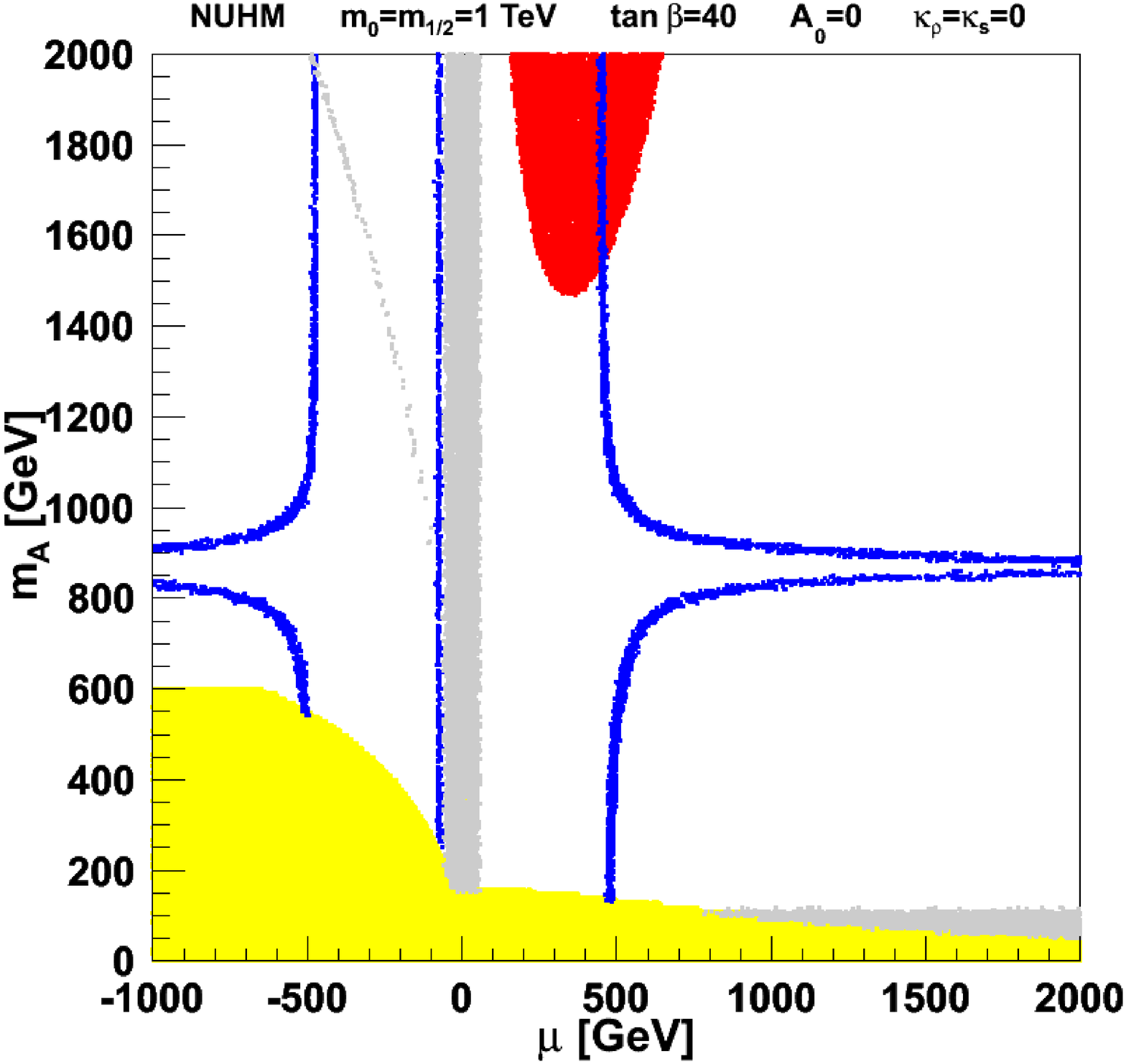}\hspace*{0.5cm}\includegraphics[width=7.4cm]{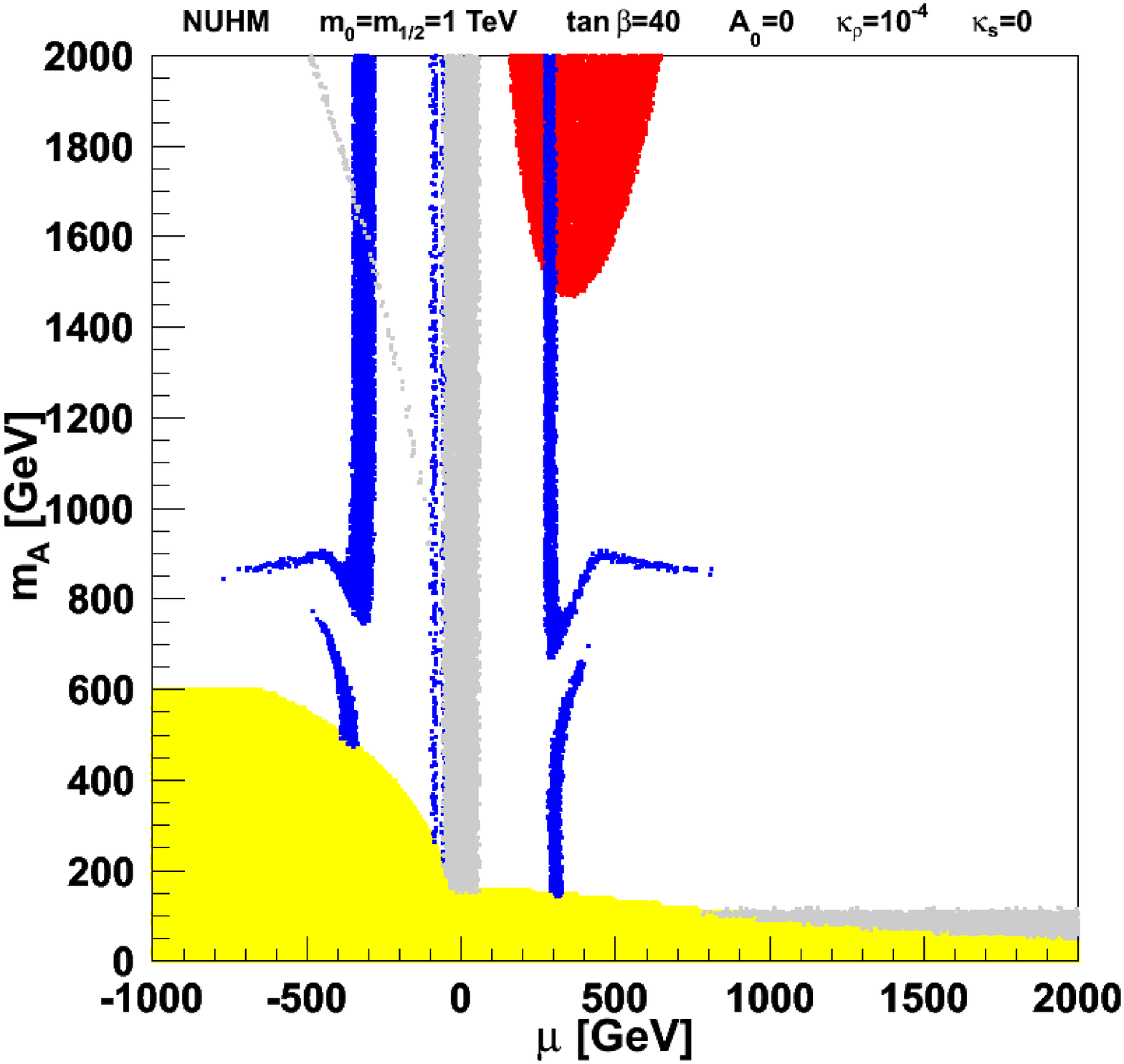}
\end{center}
\caption{Constraints on the NUHM parameter plane ($\mu$,$m_A$), in the standard cosmological model (left), and in presence of a tiny energy overdensity with $\kappa_\rho=10^{-4}$ and $n_\rho=6$ (right). The red points are excluded by the isospin asymmetry of $B \to K^* \gamma$, the gray area is excluded by direct collider limits, the yellow zone involves tachyonic particles, and the blue strips are \underline{favored} by the WMAP constraints.\label{fignuhm}}
\end{figure}

%
\verb?SuperIso Relic? computes the relic density, and the results have been compared extensively to those of \verb?DarkSusy? and \verb?Micromegas?. A very good agreement has been found even at the level of the calculation of the effective annihilation rate $W_{\rm eff}$ (see Appendix \ref{sigmav}), as can be seen in Fig. \ref{WeffDS}. In general, the results of \verb?DarkSusy?, \verb?Micromegas?, and \verb?SuperIso Relic? differ only by a few percents, but in some rare cases where a Higgs resonance occurs approximately at twice the mass of the LSP, the differences can be large. To avoid this problem, a very precise calculation of the widths of the Higgs bosons is required, and we decided to use the two-loop calculations of \verb?FeynHiggs? to obtain a better evaluation of the relic density in this kind of scenarios.\\
\\
\verb?SuperIso Relic? can also be used in order to constrain SUSY parameter spaces, as it provides many different observables from flavor physics as well as the relic density. It allows in particular to test easily the influence of the cosmological model by modifying for example the QCD equation-of-state (Appendix \ref{qcdstate}) or the expansion rate (Appendix \ref{mocomo}), as can be seen in Fig.~\ref{fignuhm}. This unusual feature will be further developed in the next versions of the program.
%
\newpage
\appendix
%
\append{Thermally averaged annihilation cross section}
\label{sigmav}
The computation of the thermally averaged annihilation cross section $\langle \sigma v \rangle$ is the most time consuming part of the relic density computation, as it requires the computation of the many annihilation and co-annihilation amplitudes. One can define the annihilation rate of supersymmetric particles $i$ and $j$ into SM particles $k$ and $l$ \cite{relic_calculation1,relic_calculation2}:
\begin{equation}
W_{ij\to kl} = \frac{p_{kl}}{16\pi^2 g_i g_j S_{kl} \sqrt{s}} \sum_{\rm{internal~d.o.f.}} \int \left| \mathcal{M}(ij\to kl) \right|^2 d\Omega \;,
\end{equation}
where $\mathcal{M}$ is the transition amplitude, $s$ is the center-of-mass energy, $g_i$ is the number of degrees of freedom of the particle $i$, $p_{kl}$ is the final center-of-mass momentum such as
\begin{equation}
p_{kl} = \frac{\left[s-(m_k+m_l)^2\right]^{1/2} \left[s-(m_k-m_l)^2\right]^{1/2}}{2\sqrt{s}}\;,
\end{equation}
$S_{kl}$ is a symmetry factor equal to 2 for identical final particles and to 1 otherwise, and the integration is over the outgoing directions of one of the final particles. Moreover, an average over initial internal degrees of freedom is performed.\\
\\
We also define an effective annihilation rate $W_{\rm eff}$ by
\begin{equation}
g_{LSP}^2 p_{\rm{eff}} W_{\rm{eff}} \equiv \sum_{ij} g_i g_j p_{ij} W_{ij}
\end{equation}
with
\begin{equation}
p_{\rm{eff}}(\sqrt{s}) = \frac{1}{2} \sqrt{(\sqrt{s})^2 -4 m_{LSP}^2} \;.
\end{equation}
Practically, we compute
\begin{equation}
\frac{d W_{\rm eff}}{d \cos\theta} = \sum_{ijkl} \frac{p_{ij} p_{kl}}{ 8 \pi g_{LSP}^2 p_{\rm eff} S_{kl} \sqrt{s} }
\sum_{\rm helicities} \left| \sum_{\rm diagrams}  \mathcal{M}(ij \to kl) \right|^2 \;,
\end{equation}
where $\theta$ is the angle between particles $i$ and $k$. We integrate over $\cos\theta$ numerically by means of the Gauss-Legendre quadrature of order 5.\\
\\
Since $W_{\rm eff}(\sqrt{s})$ does not depend on the temperature $T$, it can be tabulated once for each model point. It is however important to make sure that the maximum $\sqrt{s}$ in the table is large enough to include all important resonances, thresholds and coannihilation thresholds.\\
\\
To improve the calculation speed, we use two different thresholds:
\begin{itemize}
\item a cut such that the coannihilation of SUSY particles $i$ and $j$ is only taken into account if
\begin{equation}
 m_i + m_j < \sqrt{s}_\text{cut coann} \;,
\end{equation}
where we have taken
\begin{equation}
 \sqrt{s}_\text{cut coann} = 3 \, m_{LSP} \;,
\end{equation}
\item a maximum energy up to which $W_{\rm eff}(\sqrt{s})$ is calculated, such that
\begin{equation}
\sqrt{s}_{max}=2 m_{LSP} - T_{fo} \log(B_{\epsilon}) \;,
\end{equation}
where $T_{fo}=25$ GeV is a typical upper limit freeze-out temperature, and $B_{\epsilon}$ is the Boltzmann suppression factor limit that we fixed at $10^{-6}$ \cite{micromegas2}.
\end{itemize}
The thermal average of the effective cross section is then
\begin{equation}
\langle \sigma_{\rm{eff}}v \rangle = \dfrac{\displaystyle\int_0^\infty dp_{\rm{eff}} p_{\rm{eff}}^2 W_{\rm{eff}}(\sqrt{s}) K_1 \left(\dfrac{\sqrt{s}}{T} \right) } { m_{LSP}^4 T \left[ \displaystyle\sum_i \dfrac{g_i}{g_{LSP}} \dfrac{m_i^2}{m_1^2} K_2 \left(\dfrac{m_i}{T}\right) \right]^2}\;,
\end{equation}
where $K_1$ and $K_2$ are the modified Bessel functions of the second kind of order 1 and 2 respectively. The average is performed numerically using a Gaussian integration, and the $\infty$ limit can be safely replaced by $p_{\rm{eff}}(\sqrt{s}_{max})$ using the properties of $K_1$.
%
\append{Cosmological Standard Model}
\label{costamo}
The cosmological standard model is based on a Friedmann-Lema{\^\i}tre Universe filled with radiation, baryonic matter and cold dark matter, approximately flat and incorporating a cosmological constant accelerating its expansion. Before recombination, the Universe expansion was dominated by a radiation density, and therefore the expansion rate $H$ of the Universe is determined by the Friedmann equation
\begin{equation}
H^2=\frac{8 \pi G}{3} \rho_{rad}\;,\label{friedmann_stand}
\end{equation}
where
\begin{equation}
\rho_{rad}(T)=g_{\mbox{eff}}(T) \frac{\pi^2}{30} T^4
\end{equation}
is the radiation density and $g_{\mbox{eff}}$ is the effective number of degrees of freedom of radiation. The computation of the relic density is based on the solution of the Boltzmann evolution equation \cite{relic_calculation1,relic_calculation2}
\begin{equation}
dn/dt=-3Hn-\langle \sigma_{\mbox{eff}} v\rangle (n^2 - n_{\mbox{eq}}^2)\;, \label{evol_eq}
\end{equation}
where $n$ is the number density of all supersymmetric particles, $n_{\mbox{eq}}$ their equilibrium density, and $\langle \sigma_{\mbox{eff}} v\rangle$ is the thermal average of the annihilation rate of the supersymmetric particles to the Standard Model particles. By solving this equation, the density number of supersymmetric particles in the present Universe and consequently the relic density can be determined.\\
The ratio of the number density to the radiation entropy density, $Y(T)=n(T)/s(T)$ can be defined, where
\begin{equation}
s(T)=h_{\mbox{eff}}(T) \frac{2 \pi^2}{45} T^3 \;.
\end{equation}
$h_{\mbox{eff}}$ is the effective number of entropic degrees of freedom of radiation. Combining Eqs. (\ref{friedmann_stand}) and (\ref{evol_eq}) and defining $x=m_{\mbox{\small LSP}}/T$, the ratio of the LSP mass over temperature, yield
\begin{equation}
\frac{dY}{dx}=-\sqrt{\frac{\pi}{45 G}}\frac{g_*^{1/2} m_{\mbox{\small LSP}}}{x^2} \langle \sigma_{\mbox{eff}} v\rangle (Y^2 - Y^2_{\mbox{eq}}) \;, \label{main}
\end{equation}
with
\begin{equation}
g_*^{1/2}=\frac{h_{\mbox{eff}}}{\sqrt{g_{\mbox{eff}}}}\left(1+\frac{T}{3 h_{\mbox{eff}}}\frac{dh_{\mbox{eff}}}{dT}\right) \;.
\end{equation}
The freeze-out temperature $T_f$ is the temperature at which the LSP leaves the initial thermal equilibrium when $Y (T_f) = (1 + \delta) Y_{\mbox{eq}}(T_f)$, with $\delta \simeq 1.5$. The relic density is obtained by integrating Eq. (\ref{main}) from $x=0$ to $m_{\mbox{\small LSP}}/T_0$, where $T_0=2.726$ K is the temperature of the Universe today \cite{relic_calculation1,relic_calculation2}:
\begin{equation}
\Omega_{\mbox{\small LSP}} h^2 = \frac{m_{\mbox{\small LSP}} s(T_0) Y(T_0) h^2}{\rho_c^0} \approx 2.755\times 10^8 \frac{m_{\mbox{\small LSP}}}{1 \mbox{ GeV}} Y(T_0)\;,
\end{equation}
where $\rho_c^0$ is the critical density of the Universe, such as
\begin{equation}
H^2_0 = \frac{8 \pi G}{3} \rho_c^0 \;,
\end{equation}
$H_0$ being the Hubble constant.\\
\\
In practice, to improve the speed of the code, the freeze-out temperature $T_f$ is determined by solving the implicit equation:
\begin{eqnarray}
\frac{dY_{\mbox{eq}}}{dx} = - \sqrt{\frac{\pi}{45 G}} \frac{g_*^{1/2} m_{\mbox{\small LSP}}}{x_f^2} \langle \sigma_{\mbox{eff}} v\rangle \delta (2+\delta) Y^2_{\mbox{eq}} \;.\label{Tfo}
\end{eqnarray}
and the evolution equation (\ref{evol_eq}) is only solved from $T=T_f$ to $T_0$, with the initial condition $Y(T_f)=(1+\delta)Y_{\mbox{eq}}$. This method is known to provide results with less than a few percent error for the calculation of the relic density.
%
\append{QCD equation of state}
\label{qcdstate}
To evaluate the relic density, it is necessary to know the number of effective degrees of freedom $g_{\mbox{eff}}$ and $h_{\mbox{eff}}$ which give access to the energy and entropy densities of radiation. To compute them, one generally assumes that above the QCD phase transition critical temperature $T_c \sim 200$ MeV, the primordial plasma is weakly interacting because of asymptotic freedom, and can therefore be treated as an ideal gas.\\
\\
However, non-perturbative studies have shown that the QCD plasma departs strongly from the ideal gas behavior at high temperatures, and more realistic models have been studied in \cite{hindmarsh05}. In these models, below $T_c$ the hadronic degrees of freedom are modeled by an interacting gas of hadrons, while above $T_c$ the quarks and gluons are taken to interact and are replaced by hadronic models.  In \verb?SuperIso Relic?, the models depicted in \cite{hindmarsh05} are available, and can be selected in the routine \verb?Init_modeleff(int model_eff, struct relicparam*?\\
\verb?paramrelic)? by setting the value of \verb?model_eff? (see subsection \ref{mainroutines}). The different models are:
\begin{itemize}
\item Model A (\verb?model_eff?=1) which ignores hadrons completely.
\item Model B (\verb?model_eff?=2) which considers $T_c = 154$ MeV, and models hadrons as a gas of free mesons and hadrons, with a sharp switch to the hadronic gas at $T_{hg}=T_c$.
\item Model B2 (\verb?model_eff?=3) is a variation of model B constructed by scaling the lattice data by 0.9.
\item Model B3 (\verb?model_eff?=4) is a variation of model B constructed by scaling the lattice data by 1.1.
\item Model C (\verb?model_eff?=5) which assumes $T_c = 185.5$ MeV, and models hadrons as a gas of free mesons and hadrons, with a sharp switch to the hadronic gas at $T_{hg}=200$ MeV.
\item Old Model (\verb?model_eff?=0) is the old model with an ideal gas.
\end{itemize}
An example main program is provided as \verb?test_modeleff.c?. For more information about these models, the reader is referred to \cite{hindmarsh05}. 

\append{Modified Cosmological Model}%
\label{mocomo}
The density number of supersymmetric particles is determined by the Boltzmann equation:
\begin{equation}
\frac{dn}{dt} = - 3 H n - \langle \sigma v \rangle (n^2 - n^2_{eq}) + N_D  \;,\label{boltzmann}
\end{equation}
where $n$ is the number density of supersymmetric particles, $\langle \sigma v \rangle$ is the thermally averaged annihilation cross-section, $H$ is the Hubble parameter, $n_{eq}$ is the relic particle equilibrium number density. The term $N_D$ has been added to provide a parametrization of the non-thermal production of SUSY particles. The expansion rate $H$ is determined by the Friedmann equation:
\begin{equation}
 H^2=\frac{8 \pi G}{3} (\rho_{rad} + \rho_D)  \;,\label{friedmann}
\end{equation}
where $\rho_{rad}$ is the radiation energy density, which is considered as dominant before BBN in the standard cosmological model. Following \cite{arbey,arbey2}, $\rho_D$ is introduced as an effective dark density which parametrizes the expansion rate modification. The entropy evolution reads:
\begin{equation}
\frac{ds}{dt} = - 3 H s + \Sigma_D \label{entropy_evolution} \;,
\end{equation}
where $s$ is the total entropy density. $\Sigma_D$ parametrizes here effective entropy fluctuations due to unknown properties of the Early Universe. The radiation energy and entropy densities can be written as usual:
\begin{equation}
\rho_{rad}=g_{\rm{eff}}(T) \frac{\pi^2}{30} T^4 \;, \qquad\qquad s_{rad} = h_{\rm{eff}}(T) \frac{2\pi^2}{45} T^3 \;. \label{srad}
\end{equation}
Separating the radiation entropy density from the total entropy density, {\it i.e.} setting $s \equiv s_{rad} + s_D$ where $s_D$ is an effective entropy density, the following relation between $s_D$ and $\Sigma_D$ can be derived:
\begin{equation}
\Sigma_D = \sqrt{\frac{4 \pi^3 G}{5}} \sqrt{1 + \tilde{\rho}_D} T^2 \left[\sqrt{g_{\rm{eff}}} s_D - \frac13  \frac{h_{\rm{eff}}}{g_*^{1/2}} T \frac{ds_D}{dT}\right] \;.
\end{equation}
Following the standard relic density calculation method \cite{relic_calculation1,relic_calculation2}, we introduce $Y \equiv n/s$, and Eq. (\ref{boltzmann}) becomes
\begin{equation}
 \frac{dY}{dx}= - \frac{m_{LSP}}{x^2} \sqrt{\frac{\pi}{45 G}} g_*^{1/2} \left( \frac{1 + \tilde{s}_D}{\sqrt{1+\tilde{\rho}_D}} \right) \left[\langle \sigma v \rangle (Y^2 - Y^2_{eq}) + \frac{Y \Sigma_D - N_D}{\left(h_{\rm{eff}}(T) \frac{2\pi^2}{45} T^3\right)^2 (1+\tilde{s}_D)^2} \right] \;, \label{final}
\end{equation}
where $x=m_{LSP}/T$, $m_{LSP}$ being the mass of the relic particle, and
\begin{equation}
 \tilde{s}_D = \frac{s_D}{h_{\rm{eff}}(T) \frac{2\pi^2}{45} T^3}\;, \qquad\qquad \tilde{\rho}_D \equiv \frac{\rho_D}{g_{\rm{eff}} \frac{\pi^2}{30} T^4}\;,
\end{equation}
and
\begin{equation}
 Y_{eq} = \frac{45}{4 \pi^4 T^2} h_{\rm{eff}} \frac{1}{(1+\tilde{s}_D)} \sum_i g_i m_i^2 K_2\left(\frac{m_i}{T}\right) \;,
\end{equation}
where $i$ runs over all supersymmetric particles of mass $m_i$ and with $g_i$ degrees of freedom. Following the methods described in Appendix \ref{costamo}, the relic density can then be calculated:
\begin{equation}
 \Omega h^2 = 2.755 \times 10^8 Y_0 m_{LSP}/\mbox{GeV} \;.
\end{equation}
where $Y_0$ is the present value of $Y$. In the limit where $\rho_D = s_D = \Sigma_D = N_D = 0$, usual relations are retrieved. We should note here that $s_D$ and $\Sigma_D$ are not independent variables.\\
\\
In \verb?SuperIso Relic v2.5?, we adopt the parametrizations described in \cite{arbey,arbey2} for $\rho_D$ and $s_D$:
\begin{equation}
 \rho_D =  \kappa_\rho \rho_{rad}(T_{BBN}) \bigl(T/T_{BBN}\bigr)^{n_\rho} \label{rhoD}
\end{equation}and
\begin{equation}
 s_D =  \kappa_s s_{rad}(T_{BBN}) \bigl(T/T_{BBN}\bigr)^{n_s} \;,\label{sD}
\end{equation}
where $T_{BBN}$ stands for the BBN temperature. $\kappa_\rho$ and $\kappa_s$ are respectively the ratio of effective dark energy/entropy density over radiation energy/entropy density, and $n_\rho$ and $n_s$ are parameters describing the behavior of the densities. We refer the reader to \cite{arbey,arbey2} for detailed descriptions and discussions of these parametrizations.
%
\newpage
\section*{Acknowledgments}
\noindent The authors would like to thank Joackim Edsj\"o and Genevi\`eve B\'elanger for useful discussions, and Sven Heinemeyer and Thomas Hahn for their help with FeynHiggs and FormCalc.
%

\end{document}